# Absence of holelike Fermi surface in superconducting $K_{0.8}Fe_{1.7}Se_2$ revealed by ARPES

T. Qian[1], X.-P. Wang[1], W.-C. Jin[2], P. Zhang[1], P. Richard[1], G. Xu[1], X. Dai[1], Z. Fang[1], J.-G. Guo[1], X.-L. Chen[1], and H. Ding[1]

[1] *Beijing National Laboratory for Condensed Matter Physics, and Institute of Physics, Chinese Academy of Sciences, Beijing 100190, China*

[2] *Department of Physics, Renmin University, Beijing 100872, China*

We have performed an angle-resolved photoemission spectroscopy study of the new iron-based superconductor $K_{0.8}Fe_{1.7}Se_2$ ($T_c \sim 30$ K). Clear band dispersion is observed with the overall bandwidth renormalized by a factor of 2.5 compared to our local density approximation calculations, indicating relatively strong correlation effects. Only an electronlike band crosses the Fermi energy, forming a nearly circular Fermi surface (FS) at M ($\pi$, 0). The holelike band at $\Gamma$ sinks ~ 90 meV below the Fermi energy, with an indirect band gap of 30 meV to the bottom of the electronlike band. The observed FS topology in this superconductor favors ($\pi$, $\pi$) inter-FS scattering between the electronlike FSs at the M points, in sharp contrast with other iron-based superconductors which favor ($\pi$, 0) inter-FS scattering between holelike and electronlike FSs.

The surprising discovery of superconductivity over 30 K in $K_xFe_{2-y}Se_2$, which does not contain toxic arsenic element, has brought new excitement to the field of iron-based superconductivity [1]. The end member $KFe_2Se_2$ (0.5 doped electron per Fe) is the isostructural electron-doped counterpart of $KFe_2As_2$ (0.5 doped hole per Fe). Since $KFe_2As_2$ ($T_c \sim 3$ K) has only holelike Fermi surface (FS) sheets according to band calculations [2] and angle-resolved photoemission spectroscopy (ARPES) [3], it is possible for $KFe_2Se_2$ to have only electron FSs. This would be quite different from other iron-based superconductors with high $T_c$'s, which have both holelike and electronlike FSs quasi-nested via the ($\pi$, 0) wavevector [4-7]. Inter-FS scattering is widely believed to be important to iron-based superconductivity [8-13]. Another significant aspect of this "122" iron-chalcogenide material is that it seems to emerge from an antiferromagnetic (AFM) insulating phase instead of a metallic spin-density-wave (SDW) parent state as in the cases of many other iron-based superconductors. A similar superconductor, $(Tl,K)Fe_{2-y}Se_2$, was recently reported to have a $T_c$ as high as 40 K [14] when y ≤ 0.12, while becoming an AFM insulator when y ≥ 0.4. The insulating phase for the possible parent compound $(Tl,K)Fe_{1.5}Se_2$, as in the high-$T_c$ cuprates, makes this material unique and important to study.

In this letter, we present the first ARPES results on the band structure and the FS of superconducting $K_{0.8}Fe_{1.7}Se_2$ ($T_c \sim 30$ K) single crystals, and compare them to our local density approximation (LDA) band results. Our main finding is that hole bands at Γ never cross the Fermi energy ($E_F$), while an electron band crosses $E_F$ and forms a nearly circular FS at M ($\pi$, 0). In addition, an indirect band gap between the hole band and the electron band is observed. The observed bandwidth and Fermi velocity are renormalized by a factor of 2.5 compared to our LDA results, indicating a relatively strong correlation effect in this material. The implications of the distinct FS topology observed in this iron chalcogenide on its magnetic and superconducting properties will be discussed.

The single crystals of $K_{0.8}Fe_{1.7}Se_2$ (nominal concentration $K_{0.8}Fe_2Se_2$) used in this study were grown by the flux method, as described in Ref. [1]. Transport measurements indicate a superconducting onset temperature at 30 K and a transition width of 3 K. ARPES measurements were performed at the Institute of Physics, Chinese Academy of Sciences, using the He Iα ($hv$ = 21.218 eV) and He IIα ($hv$ = 40.814 eV) resonance lines. The angular resolution was set to 0.2°, while the energy resolution was set to 35 and 70 meV for the He Iα and He IIα measurements, respectively. Samples with a typical size of ~ 1×1 mm$^2$ were cleaved *in situ* and measured at 40 K in a working vacuum better than 5×10$^{-11}$ The $E_F$ of the samples was referenced to that of a gold film evaporated onto the sample holder. We start with a wide energy spectrum (Fig. 1a) that includes shallow core levels and the valence band using 40.814 eV photons. The strong peak at the binding energy of 17.55 eV is from K 3*p*, slightly smaller than the one observed in Ba$_{0.6}$K$_{0.4}$Fe$_2$As$_2$ [15]. In addition, there is a peak at 12 eV with an unknown origin. Fig. 1c displays energy distribution curves (EDCs) along several high symmetry directions (Γ-M, M-X, Γ-X), as defined in Fig. 1b. One distinct feature in this material is a large, broad and weakly dispersive peak located at 0.9 eV. While it could come from the bottom of the Fe 3*d* orbitals, as shown in Fig. 1d, it corresponds more likely to the large incoherent component of the Fe 3*d* orbitals. In fact, a LDA+DMFT calculation on FeSe has found a lower Hubbard band around 1 eV due to strong correlation effects in iron selenide materials [16]. As with other iron-based superconductors, electronic correlations lead to a renormalization of the band structure. As shown in Fig. 1d, our LDA band calculations capture several features emphasized by the intensity plot of the second derivative of EDCs along high symmetry directions. Our calculated bands have been shifted up by 170 meV to be consistent with the experimental electron doping, and then renormalized by a factor of 2.5. Both the calculations (at $k_z = 0$) and the experimental data suggest that an electronlike FS pocket emerges at the M point, defined as (π, 0) in the unfolded Brillouin zone (BZ). However, there exist obvious discrepancies between theory and experiment. The main one is the opposite energy shift between the hole and electron bands with respect to LDA results, namely a downward shift for the hole band and an upward shift for the electron band. Such electron-hole asymmetric shift has been demonstrated to be an evidence for a dominant interband coupling [17].

To investigate further the FS topology of this new superconductor, we next focus on the electronic band structure in the vicinity of the Fermi level. We display in Fig. 2a the ARPES intensity plot of a cut passing through the M point. The intensity plot clearly shows that an electronlike band crosses the Fermi level. This is further supported by its second derivative along the energy direction, given in Fig. 2b, as well as by the corresponding EDCs and momentum distribution curves (MDCs), which are displayed in Fig. 2c and Fig. 2d, respectively. From the EDCs and the MDCs, we estimate the bottom of the band at 60 meV. Taking into account this value, a simple parabolic fit allows us to estimate a Fermi velocity of 0.52 eV·Å and an electron mass of 3.5$m_0$. In addition to the electron band at the M point, the data indicate the presence of a holelike band feature topping at M around 130 meV.

In contrast to the M point, we do not see any band crossing the Fermi level at the Γ point. This is well illustrated by the ARPES intensity plot shown in Fig. 3a. Instead, both the corresponding the EDCs and second derivative intensity plot shown in Fig. 3b and Fig. 3c, respectively, indicate a holelike band topping at 90 meV. Interestingly, this value corresponds to a higher binding energy than the bottom of the electronlike band at the M point, indicating an indirect band gap of 30 meV in the band structure. We speculate that this band gap might be related to the presence of an insulating phase at lower electron doping [14]. To check whether the top of the holelike band at the Γ point can cross $E_F$ at a different $k_z$ value, we contrast the He Iα (21.218 eV) data with data recorded with the He IIα (40.814 eV) line (Fig. 3d), which corresponds to a different $k_z$. The data are quite similar, except for a slight shift of the top of band towards higher binding energies. We also plot in Fig. 3e the data obtained with the He IIα line at the second Γ, for which $k_z$ also varies. Once more, the holelike band does not cross the Fermi level. Using the conversion equation $k_z = \sqrt{2m[(h\nu - \phi - E_B)\cos^2\theta + V_0]}/\hbar$, where the inner potential $V_0$ is estimated to be about 15 eV in pnictides [18], we estimate the $k_z$ values to be about 3.2, 4.1, and 3.7 (4π/c) for Figs. 3c-e, respectively.

The ARPES intensity map integrated in the ± 20 meV energy range is given in Fig. 4a. The high intensity regions define the Fermi surface. As explained above, while one electron FS pocket is detected at the M point, there is no FS pocket observed at the BZ center. Although our experimental resolution does not allow us to resolve two electronlike FS pockets at the M point, all band calculations on iron-based superconductors, as well as previous ARPES measurements, suggest that there should be two. As a first approximation, it is thus a reasonable assumption to consider that there are indeed two electronlike FS pockets at M and that they are almost degenerate. The size of one electronlike FS pocket is estimated at 5.5% of the folded BZ. Assuming a double degeneracy, this leads to an electron concentration of 11% per Fe, in good agreement with the chemical formula.

The observed FS topology in $K_{0.8}Fe_{1.7}Se_2$ is apparently in direct contradiction with the scenario promoting (π, 0) AFM scattering between Γ-centered holelike and M-centered electronlike FS pockets as the key ingredient for Cooper pairing in iron-based superconductors. However, we point out that even though (π, 0) AFM scattering fails to explain superconductivity as high as 30 K in the absence of holelike FS pocket at the zone center, short range AFM inter-band scattering remains possible, albeit with a different wavevector. In fact, this can be viewed from Fig. 4b where the circular electronlike FSs at M are connected by the ($\pi$, $\pi$) wavevector in the unfolded BZ. The inequivalent Se sites fold the BZ with respect to the same ($\pi$, $\pi$) wavevector. Consequently, there are two bands almost degenerated at the M point, which should have different orbital characters. While a strong spin coupling constant between second iron neighbors ($J_2$) favors ($\pi$, 0) AFM scattering, G-type AFM ordering and ($\pi$, $\pi$) AFM scattering are favored by a strong coupling between nearest iron neighbors ($J_1$). The latter scenario is compatible with the observed FS topology. We caution that the current ARPES results do not allow us to distinguish between phonon-driven intraband scattering and AFM interband scattering. However, different superconducting pairing symmetries are expected for these two scenarios. Interestingly, Kuroki *et al.* predicted that although a s$_{+/-}$ paring symmetry is expected when both Γ-centered holelike and M-centered electronlike FS pockets are present, the absence of holelike FS pocket at Γ would favor a *$dx^2$-$y^2$* pairing symmetry [11].

In conclusion, our ARPES results show that while an electron band crosses $E_F$ and forms a nearly circular FS at the M ($\pi$, 0) point, a hole band at the Γ point never crosses $E_F$. The Luttinger volume of electron FSs is ~ 11% of BZ area assuming a double degeneracy, consistent with the valence counting. An indirect band gap of ~ 30 meV between the top of the hole band and the bottom of the electron band is also observed, suggesting an insulating state at lower electron doping levels. Our LDA results on KFe$_2$Se$_2$ capture many dispersive features observed by ARPES when normalized by a factor of 2.5, indicating a relatively strong correlation effect in this material. Unlike many other iron-based superconductors where ($\pi$, 0) scattering between holelike FSs at Γ and electronlike FSs at M is believed to dominate, this iron-based superconductor likely favors ($\pi$, $\pi$) inter-FS scattering between the electronlike FSs at M.

This work was supported by grants from Chinese Academy of Sciences, Ministry of Science and Technology of China and Chinese National Science Foundation.

**Fig. 1** (a) Photoemission spectra recorded with the He IIα resonance line ($h\nu$ = 40.814 eV). (b) Schematic definition of the Γ(0, 0), M(π, 0) and X(π/2, π/2) high symmetry points. Our notation refers to the unfolded Brillouin zone, which corresponds to the 1 Fe site/unit cell description. (c) EDCs along several high symmetry directions recorded with the He Iα resonance line ($h\nu$ = 21.218 eV). Blue curves correspond to high symmetry points. (d) Second derivative intensity plot along high symmetry lines. The experimental data are compared to our LDA band structure calculations on $KFe_2Se_2$ ($k_z$ = 0), which have been shifted up by 170 meV to account for the electron doping and then renormalized by a factor 2.5.

**Fig. 2** (a) ARPES intensity plot along a cut passing through M ($h\nu$ = 21.218 eV). The red dashed curve is a parabolic. (b) Corresponding second derivative intensity plot. EDCs and MDCs along the cut shown in (a) are given in (c) and (d), respectively.

**Fig. 3** (a) ARPES intensity plot recorded with the He Iα resonance line ($h\nu$ = 21.218 eV) along a cut passing through Γ ($k_z$ = 12.9 π/c). (b) EDCs along the cut shown in (a). (c) Second derivative intensity plot of the cut shown in (a). (d) and (e) correspond to second energy derivative intensity plot recorded with the He IIα resonance line ($h\nu$ = 40.814 eV) along a cut passing through Γ ($k_z$ = 16.4 π/c) and second Γ (Γ'; $k_z$ = 14.8 π/c), respectively.

**Fig. 4** (a) ARPES intensity mapping recorded with $h\nu$ = 21.218 eV photons and integrated within ± 20 meV. (b) Schematic diagram summarizing the electronic band structure of $K_{0.8}Fe_{1.7}Se_2$ and illustrating the (π, π) scattering processes.

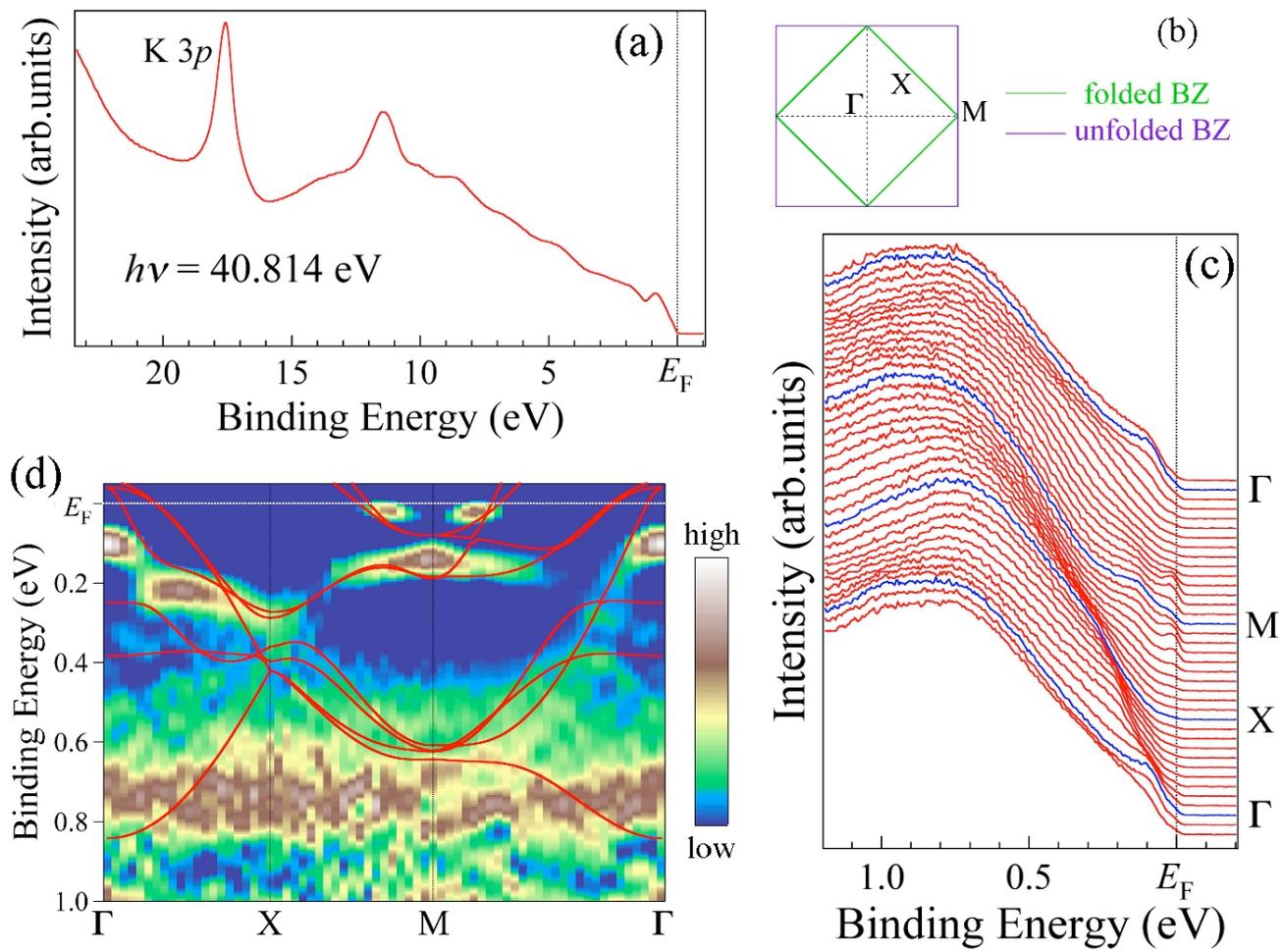

Fig. 1

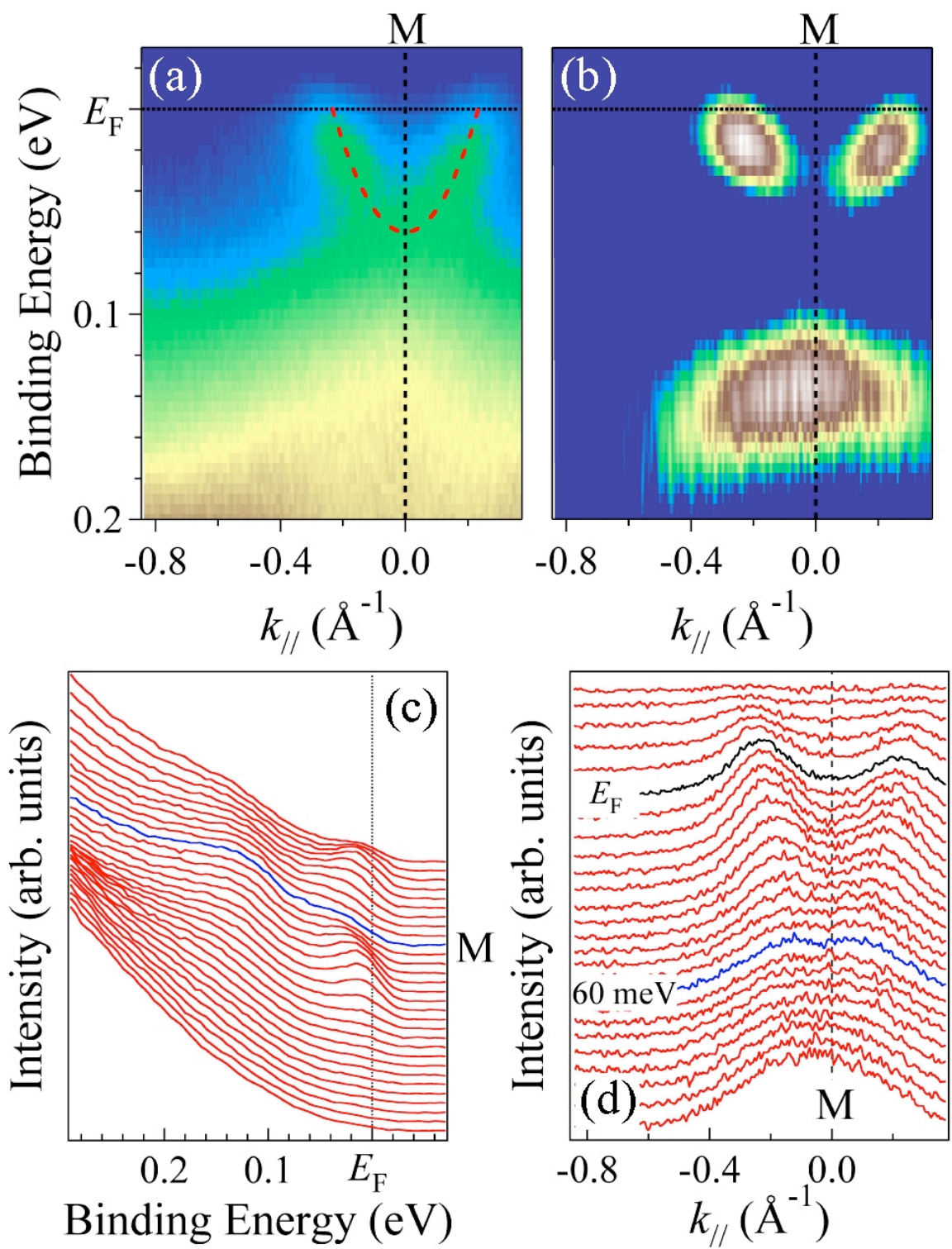

Fig. 2

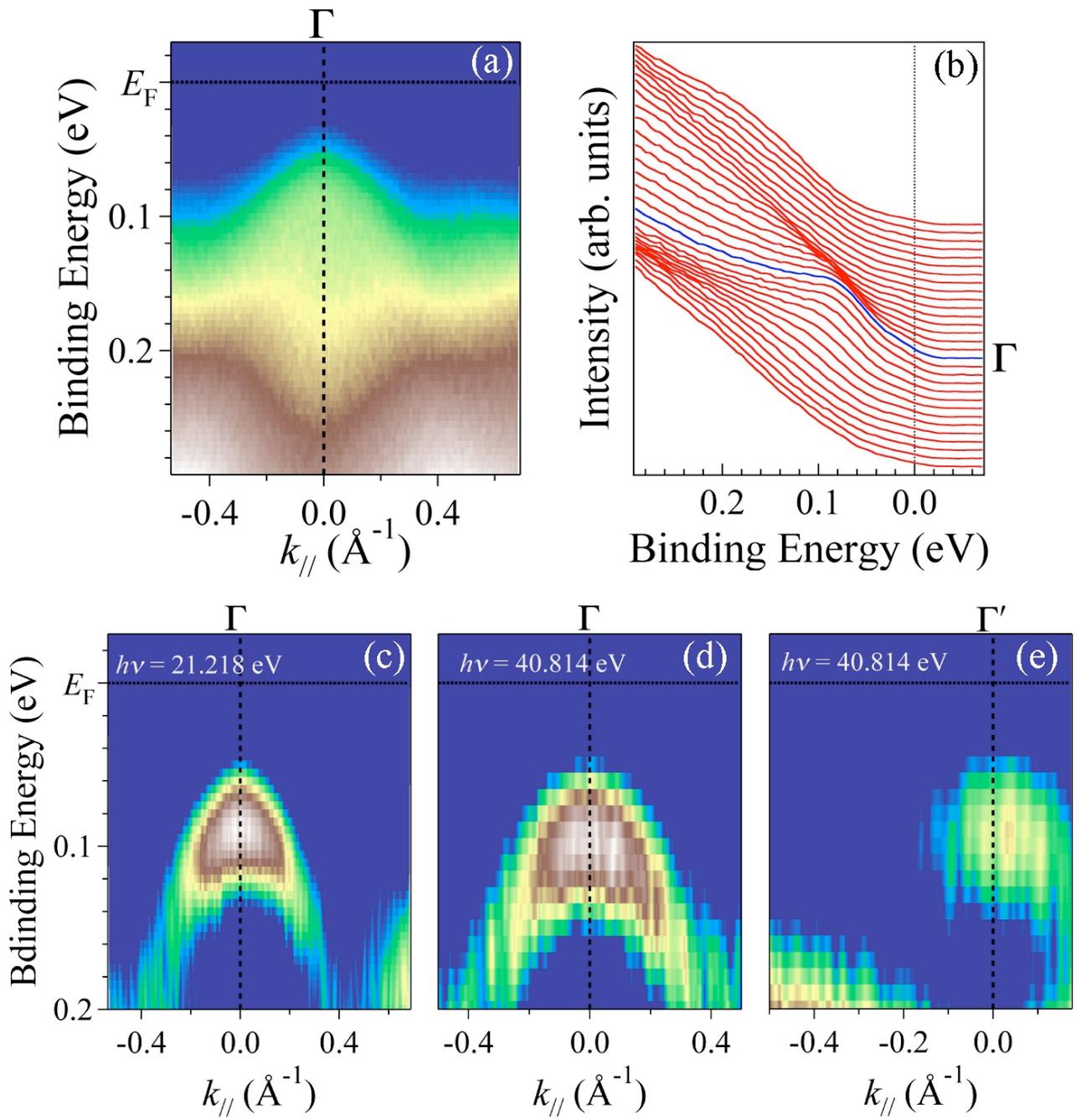

Fig. 3

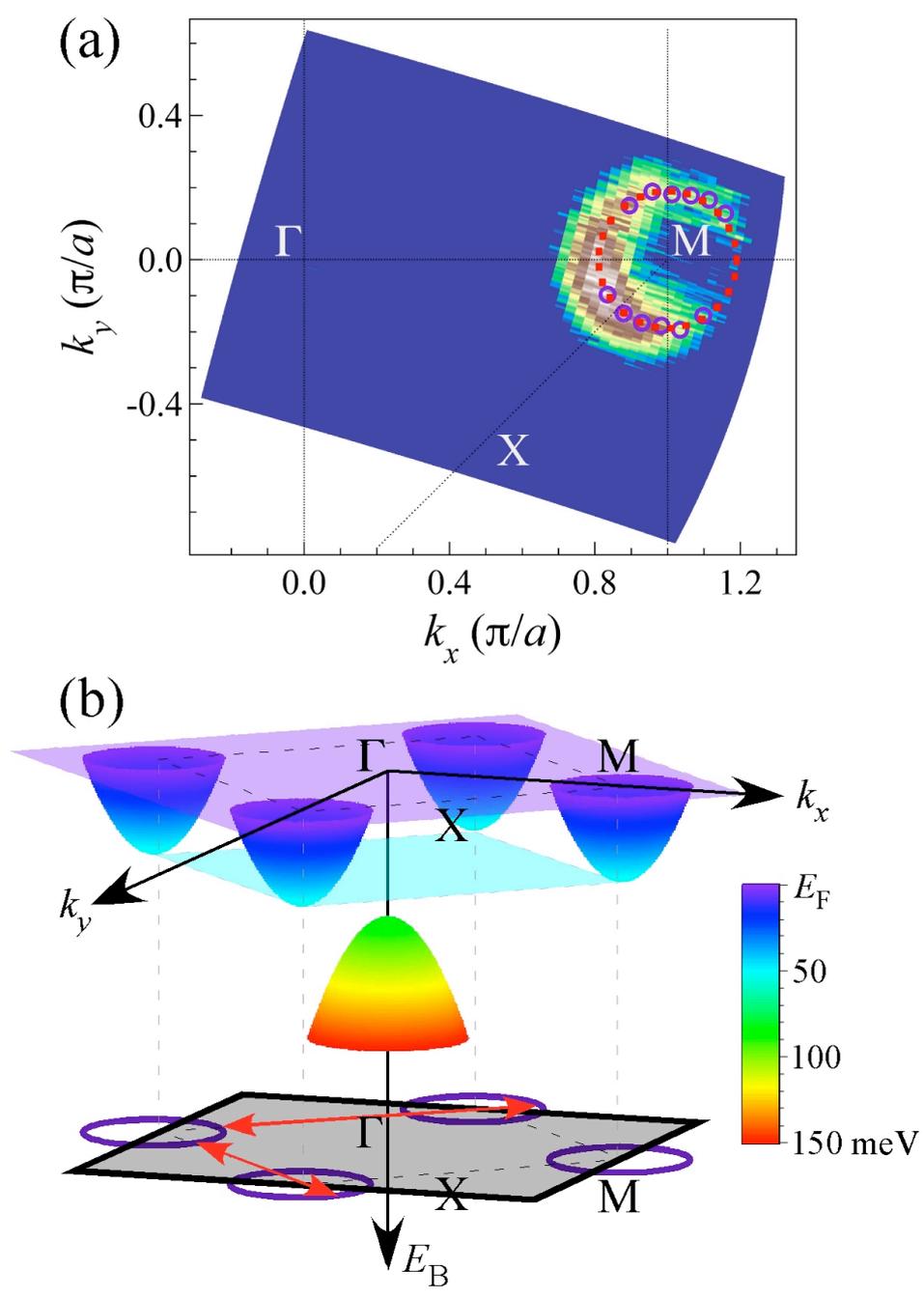

Fig. 4